\documentstyle[12pt,twoside,fleqn,espcrc1,epsf]{article}

\newcommand{\be}{\begin{equation}} 
\newcommand{\en}{\end{equation}}
\newcommand{\bea}{\begin{eqnarray}}
\newcommand{\ena}{\end{eqnarray}}
\newcommand{\hbo}{\hbox to 1 true cm {\hfill } }

\def\dslash{\partial\kern-.5em\slash}

\newcommand{\AmS}{{\protect\the\textfont2
  A\kern-.1667em\lower.5ex\hbox{M}\kern-.125emS}}

\title{ Induced symmetry breaking and a new phase of hadronic matter 
       at high density }

\author{K.~Langfeld\address{Institut f\"ur Theoretische Physik, 
        Universit\"at T\"ubingen, \\ 
        Auf der Morgenstelle 14, D-72076 T\"ubingen }%
}

\begin{document}
\maketitle

\begin{abstract}
The notion {\it induced symmetry breaking } (ISB) is introduced as a 
generalization of the spontaneous symmetry breaking mechanism and is 
illustrated in a simple two flavor spin model. 
In the case of QCD at finite baryon density, I argue that the quark 
interaction induced by {\it zero sound} satisfies the prerequisites 
which are necessary for a ISB scenario. 
In this scenario, the quark condensate sharply drops at the 
critical value of the chemical potential in coincidence with a rapid 
increase of the baryon density. The spectrum of the light particles is 
discussed below and above this phase transition. The consequences of 
the ISB mechanism for heavy ion collisions are briefly addressed.

\end{abstract}

\section{Introduction}

The understanding of low energy QCD requires powerful non-perturbative 
techniques. Such a technique was already outlined by Goldstone 
in the early sixties~\cite{gol61,yn83}. He pointed out that the vacuum 
structure does not necessarily respect the symmetries of the Lagrangian. 
In this event, 
the symmetry is spontaneously broken and massless excitations exist which 
carry the quantum numbers of the generators corresponding to the 
spontaneously broken. In particular, assuming that the chiral symmetry 
of QCD is spontaneously broken, one understands the particular role 
of the light pseudo-scalar mesons in the spectrum without solving 
the full strong coupling problem. 

\vskip0.3cm 
The efforts to understand the evolution of the 
universe~\cite{kol90} or compact star matter~\cite{rho1} has posed 
a new challenge, the importance of which has steadily grown 
in the last years: the properties of hadrons at intermediate and high 
baryon density. This subject has attracted recent interest due to 
the data of the CERES and the HELIOS groups, who reported deviations of 
heavy ion collision data from the theoretical predictions according to 
present understanding~\cite{ceres,helios}. They find that the dilepton 
yield in Au-Au collisions is enhanced in the invariant mass range of 
$300 \ldots 600 \, $MeV compared with the collisions p--Be and 
p--Au~\cite{ceres,helios}. The experimental data 
have launched a vivid discussion on the origin of the dilepton enhancement. 
Most proposals focus on the role of the in medium $\rho $-meson, which 
dominantly couples to photons. One finds a good agreement of theory and 
experiment in S-Au collisions, if a in medium $\rho $ 
mass-shift~\cite{rho91} towards smaller values is included in the 
relativistic transport calculations~\cite{ko96,ko96b}. 
It was subsequently observed that 
a broadening of the $\rho $-width in baryonic matter also results 
in a dilepton enhancement compatible with the accuracy of the present 
experimental data~\cite{wam96,son96,kli97}. Both explanations are 
so far compatible with the prediction of QCD sum rules~\cite{hat92,kli97}. 

\vskip0.3cm 
In this paper, I will employ the powerful concept of symmetry breaking 
to study QCD baryon matter. The notion of {\it induced symmetry breaking} 
(ISB) is introduced with the help of a simple two flavor spin model. 
I will discuss the prerequisites which are necessary for the ISB 
scenario to occur in the case of QCD at finite chemical potential, 
and will argue that an effective quark theory the interactions of which 
are mediated by {\it zero sound } satisfies these conditions. 
In this case, a new state of hadronic matter occurs above a critical 
value of the chemical potential. In this new phase, a light vector 
meson emerges which dominantly couples to dileptons. 
The new state of matter might play a role for the dilepton 
enhancement actually observed in the CERES and HELIOS experiments.

\section{ The concept of induced symmetry breaking} 

\vskip 0.3cm 
Let us study the basic features of the ISB scenario in a simple 
spin model from solid state physics. For this purpose, 
we will briefly study a liquid consisting of two (classical) spins, one 
of unit length (''slow'' degree of freedom), $\vec{s}_l, \; 
\vec{s}_l^2 =1$, and the other spin $\vec{s}_f$ with fixed length 
$\vert s_f \vert <1 $ and high mobility (''fast'' degree of freedom). 
The two different types of spins possess a quadrupol type interaction 
implying that the screening and anti-screening, respectively, of the 
''slow'' degrees of freedom by the ''fast'' ones is not biased 
(for zero magnetic field). The ''slow'' degree of freedom $\vec{s}_l$ is 
dressed by a cloud of spins $\vec{s}_f$. For an {\it effective} 
description of the ''slow'' degrees of freedom, it is therefore convenient 
to introduce the notion of a constituent spin $\vec{\phi }$. 
$\vec{\phi }$ is a collective degree of freedom composed of $\vec{s}_l$ 
and $\vec{s}_f$, averaged over $\vec{s}_f$. Fluctuations of the 
$\vec{s}_f$-cloud introduces a variable length for the constituent 
spin $\vec{\phi }$. Within the framework of this 
effective approach to the two-flavor spin liquid, the 
partition function, 
\be 
Z \; = \; \int [d \Omega _x] \; \exp \left\{ - \beta H \right\} \; , 
\label{eq:p1} 
\en 
is a functional integral over the angular degrees of freedom $\Omega _x$ 
of $\vec{s}_l$. In order to be precise, I confine myself to 
3-dimensional vectors. I propose the following effective 
Hamiltonian, 
\be 
H \; =\;  - \sum _{\langle xy \rangle } \vec{\phi }_x \vec{\phi }_y 
\; + \; V(\vec{\phi }^2) \; - \; \sum _x \vec{\phi }_x \vec{B} \; , 
\hbo 
V(\vec{\phi }^2)  \; = \; \lambda \, \vec{\phi }^2 \, ( \vec{\phi }^2 \, - \, 
2) \; , 
\label{eq:4.71} 
\en 
where the ''constituent'' spins $\vec{\phi }$ are located at the lattice 
sites $x$, and where the sum over 
$\langle xy \rangle$ extends over nearest neighbors. 
The effective length of the ''constituent'' spins $\vec{\phi }$ is 
mainly controlled by the potential $V(\vec{\phi }^2)$ and can be 
obtained from $ \langle H \rangle _{\Omega } \rightarrow \hbox{minimum} $. 
The potential $V(\vec{\phi }^2)$ 
possesses its global minimum at $\vec{\phi }^2=1$, where 
the curvature of the potential at the minimum is governed by $\lambda $. 
The fields $\vec{\phi }_x$ interact with a constant external ''magnetic'' 
field $\vec{B}$. We assume that $\lambda $ is independent of the 
magnetic field in the range of interest. 
In the limit $\vec{B}=0$, the model defined by (\ref{eq:4.71}) 
is invariant under a global rotation of the vectors. 

\begin{figure}[htb]
\begin{minipage}[t]{80mm}
\epsfxsize=7cm
\epsffile{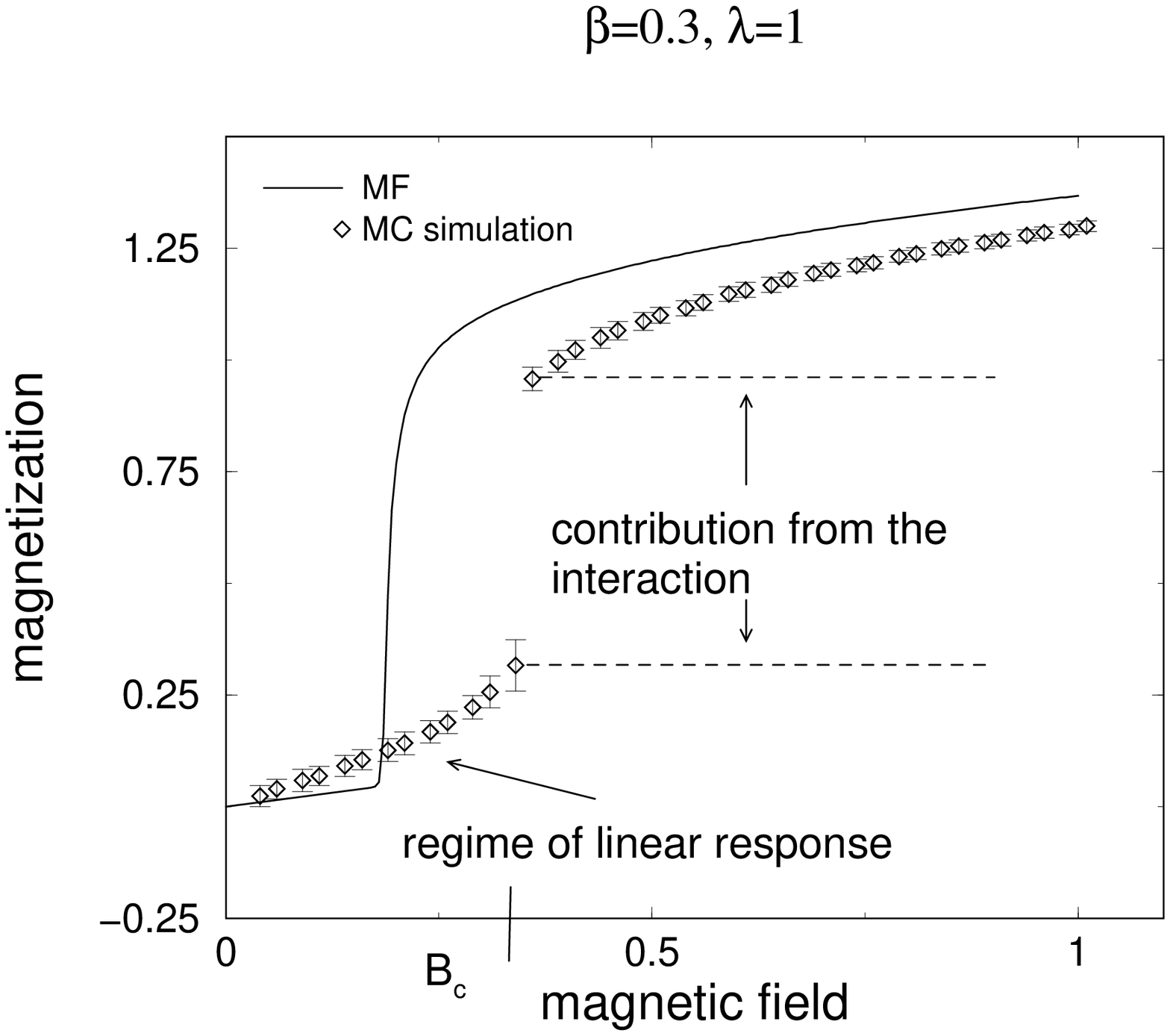} 
\caption{ The magnetization as function of the external magnetic field 
   for an inverse temperature $\beta = 1/T = 0.3 < 1/T_{Curie}$. } 
\label{fig:1}
\end{minipage}
\hspace{\fill}
\begin{minipage}[t]{75mm}
\epsfxsize=6cm
\epsffile{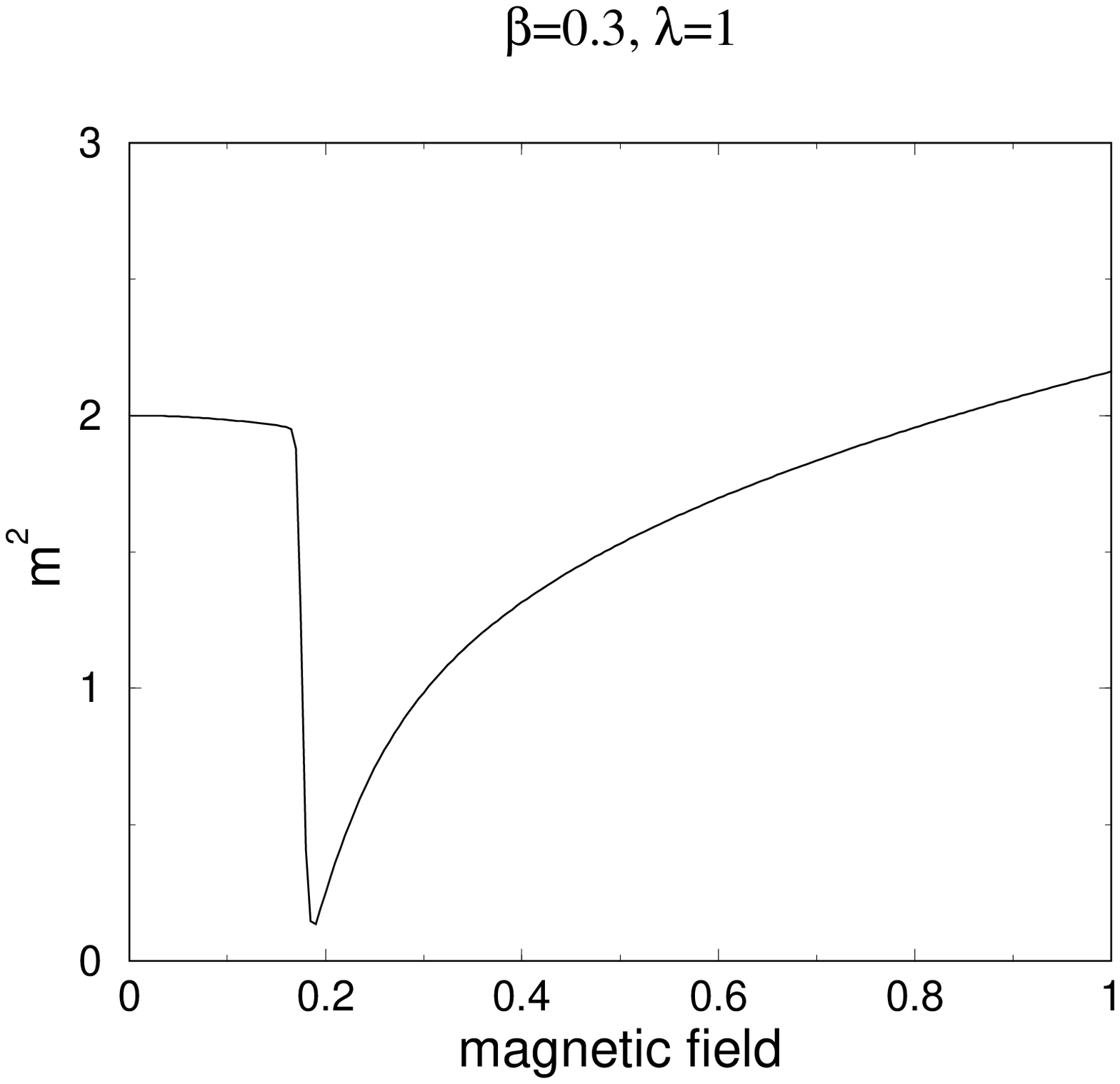} 
\caption{ The behavior of the screening mass as function of the external 
   magnetic field (MF-approximation). } 
\label{fig:2}
\end{minipage}
\end{figure}

The theory (\ref{eq:4.71}) of the constituent spin can be easily 
studied employing numerical simulations. It turns out that the mean 
field approach reproduces the gross features. For $\lambda 
\gg 1 $, the model (\ref{eq:4.71}) is reduced to an Heisenberg spin model. 
In this case and for temperatures larger than the Curie temperature, 
the magnetization monotonically increases with the external magnetic 
field. For $\lambda \approx 1$, the spin 
system shows the ISB behavior: for values of the magnetic field smaller 
than the critical value $B_c$, one observes a linear response 
of the magnetization with increasing external field. At the critical 
value $B_c$, a rapid increase of the magnetization takes place (see 
figure \ref{fig:1}). 

\vskip 0.3cm 
This phenomenon of ISB can be understood as follows: since the 
interaction of ''slow'' and ''fast'' degrees of freedom does not prefer 
screening or anti-screening, the average length of the constituent 
spin is of order unity at zero magnetic field. At non-zero magnetic 
field, both $\vec{s}_l$ and $\vec{s}_f$ are partially aligned along 
the direction of the magnetic field and the length of the constituent 
spins is shifted towards larger values on average. Finally, the 
residual interaction of the constituent spins is large enough for 
a spontaneous breaking of rotational symmetry and {\it induced 
symmetry breaking } takes place. 

\vskip 0.3cm 
Since the interaction provides a large contribution to the 
order parameter, i.e. the magnetization, in addition to the 
(small) contribution from the external magnetic field, one expects 
the presence of long range correlations by virtue of Goldstone's theorem. 
In order to reveal these correlations, it is convenient 
to define the screening mass $m$ by 
\be 
\frac{1}{m^2} \delta _{ik} \; = \; \int d^3x \; \langle 
\phi _i (\vec{x} ) \, \phi _k (0) \, \rangle \; = \; 
\frac{ \partial {\cal M}_i} { \partial B_k } \; , 
\en 
where ${\cal M}_i := \langle \phi _i \rangle $ is the magnetization 
and $\vec{B}$ is the {\it constant } external magnetic field. 
Figure \ref{fig:2} shows the dependence of the screening mass 
on the external magnetic field. At the critical value $B_c$, one 
observes a sudden decrease of $m$ corresponding to long range 
correlations. Note that the excitations are not strictly massless 
due to the explicit breaking of rotational symmetry by the 
external field $\vec{B}$. For values $B>B_c$, the screening mass 
increases. Finally, the concept of an approximate symmetry becomes 
meaningless for a strong explicit symmetry breaking, i.e. $B \gg B_c$. 

\section{ ISB and QCD at finite chemical potential } 

\subsection{ Prerequisites for ISB } 

\begin{figure}[htb]
\begin{minipage}[t]{80mm}
\epsfxsize=7cm
\epsffile{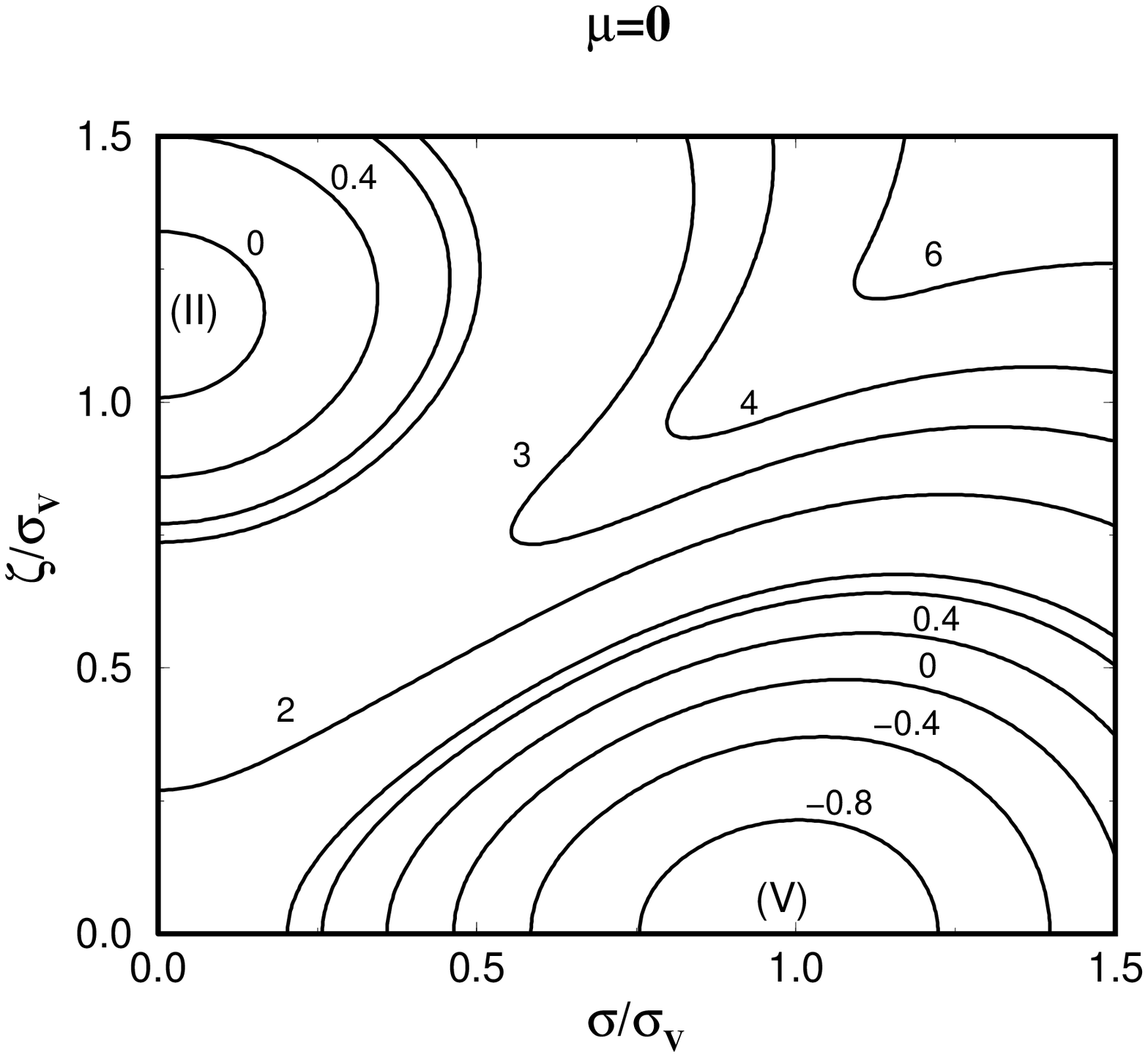} 
\caption{ Lines of constant effective potential as function of the 
   quark condensate $\sigma $ and the baryon density $\zeta $ at 
   zero chemical potential. } 
\label{fig:3}
\end{minipage}
\hspace{\fill}
\begin{minipage}[t]{75mm}
\epsfxsize=7cm
\epsffile{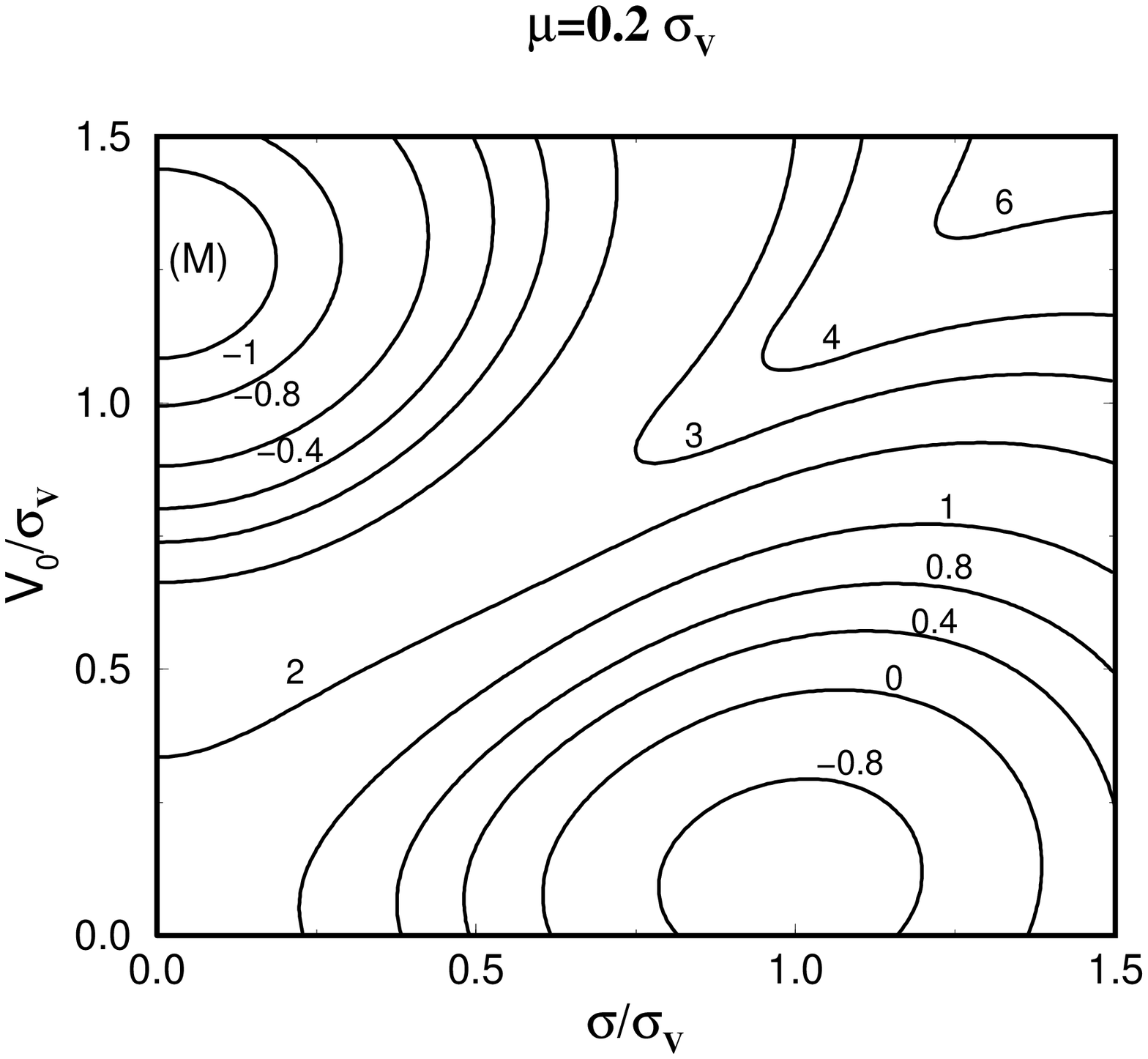} 
\caption{ Same as figure 3 at a finite value of the chemical potential. } 
\label{fig:4}
\end{minipage}
\end{figure}
\begin{figure}[htb]
\epsfxsize=12cm
\epsffile{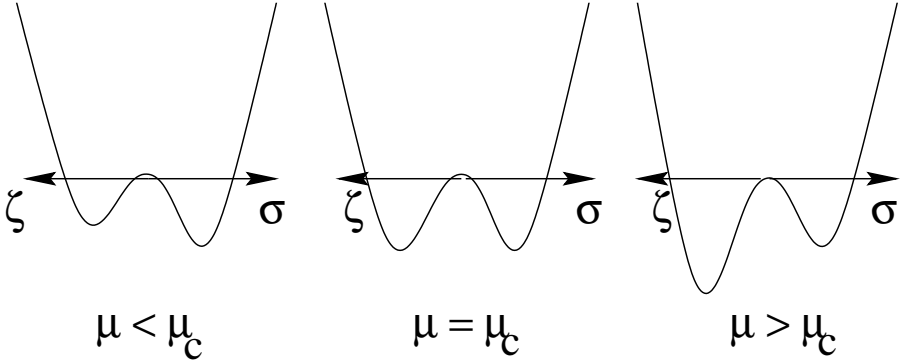} 
\caption{ Schematic plot of the effective potential at the axis 
   $(\sigma , 0)$ and $(0,\zeta )$ for several values of the 
   chemical potential $\mu $. } 
\label{fig:5}
\end{figure}

The sole condition for ISB to occur in hadron matter~\cite{larr97} can be 
most easily anticipated from the effective potential as function of the 
quark condensate $\sigma =\langle \bar{q}q \rangle $ and the 
baryon density $\zeta =\langle \bar{q} \gamma _0q \rangle $ as sketched 
in figure \ref{fig:3}. Whereas the {\it global} minimum at $(\sigma , \zeta ) 
=(\sigma _v,0)$ constitutes the vacuum, the crucial assumption is that 
a second, {\it local } minimum, i.e. $II$ in figure \ref{fig:3}, 
occurs at $(0,\zeta \not=0)$. At finite values of the chemical potential 
$\mu $, this second minimum $II$ gets more pronounced than the state 
$V$, since $\mu $ acts as a Lagrange multiplier for the baryon density, 
whereas the impact of $\mu $ on the quark condensate is of indirect 
nature. Increasing the chemical potential $\mu $, the generic 
behavior of the effective potential (at the axis) is illustrated 
in figure \ref{fig:5}. 
At $\mu < \mu _c$, the non-vanishing values of the quark condensate 
signals a spontaneous breakdown of chiral symmetry. For $\mu >\mu _c$, 
the state $II$ becomes the true ground state. Chiral symmetry is 
restored ($\sigma \approx 0$), and the interaction strongly contributes 
to baryon density $\zeta $. ISB as discussed in the previous section 
sets in. 

\vskip 0.3cm 
The change of the state structure at $\mu = \mu _{ci}$ tremendously 
influences the particle spectrum: firstly, we expect from Goldstone's 
theorem the presence  of a light particle which carries the quantum 
numbers of a flavor singlet (non-relativistic) vector meson~\cite{larr97}. 
On the other hand, chiral symmetry is restored, and the pion ceases to 
play a special role as the lightest hadron. We therefore expect its mass 
in the range of several hundred MeVs. The properties of the emerging vector 
particle were studied in~\cite{larr97} in great detail. In particular, one 
expects from Goldstone's theorem that this vector particle dominantly 
couples to dilepton pairs. We were able to derive a Gell-Mann-Oakes-Renner 
type of relation~\cite{larr97}. Our formula 
\be 
m_V^2 f_V ^2 \; = \;  2 \mu \langle \bar{q} \gamma _0 q \rangle \; , 
\label{eq:4.66} 
\en 
relates the mass of the vector particle $m_V$ and its decay 
constant $f_V$ to the chemical potential $\mu $ and the baryon 
density. Relation (\ref{eq:4.66}) quantifies the influence of the 
small explicit (on top of the spontaneous) breaking of Lorenz invariance 
via the chemical potential. 

\subsection{ Quark interactions mediated by zero sound } 

The above scenario at finite density does not invoke any particular 
quark model, but relies on the existence of the local minimum $II$. 
Whether this minimum exists in the case of QCD is up to now an 
open question. In the following, I will argue that the quark interactions 
induced by zero sound may meet the prerequisites for an ISB scenario. 
For this purpose, we studied the following effective quark 
theory~\cite{larr97} 
\bea 
Z[s,j_\mu] &=& \int {\cal D} q \, {\cal D} \bar{q} \, {\cal D} \sigma \, 
{\cal D} \pi \; {\cal D} V_\mu \; e^{ 
- \int d^4x \; [L - s(x) \sigma (x) - j_\mu (x) V^\mu (x) ]} \; , 
\label{eq:4.64} \\ 
L &=& \bar{q}(x) \Big( i \dslash -  \sigma (x) + i \gamma _5 \pi (x) 
+ i V_\mu (x) \gamma ^\mu \Big) q(x) 
\label{eq:4.65} \\ 
&+& \frac{N}{2} g_s \left[(\sigma (x)-m)^2 + \pi ^2(x) \right] 
\nonumber \\ 
&+& \frac{N}{2} \left\{ V_\mu (x) \left( - \partial ^2 \delta _{\mu \nu } 
+ \partial _\mu \partial _\nu \right) V_\nu (x) 
\; + \; m^2_v \left[ (V_0(x) - \mu )^2 + V_k^2(x) \right] \right\} \; , 
\nonumber 
\ena 
where $N$ is the number of colors. $\sigma $ and $\pi $ denote 
the usual scalar and pionic hadron fields. In addition to these fields, 
a vector--like quark interaction is mediated by the field $V_\mu (x)$.
Note that the coupling of these vector fields to the quark is 
defined in such a way that the zero component $V_0$ acts like 
a chemical potential. Note that this is different from the standard 
coupling of $\rho $ and $\omega $ mesons to quarks. 
Fluctuations of $V_0$ act like fluctuations in the chemical potential. 
At the level of the effective quark theory (\ref{eq:4.64}), the vector 
field $V_\mu $  therefore describes a vector type quark interaction due 
to density fluctuations, which is present on top of the 
interactions mediated by the mesons $\sigma $, $\pi $, $\rho $, 
$\omega $. 
The introduction of the vector field $V_\mu $ is reminiscent of the 
solid state physics phenomenon of {\it zero sound}~\cite{sound}, which 
can be observed in Fermi liquids.   
Whereas ''first sound'' describes density fluctuations which are due 
to distortions of the thermodynamical equilibrium, 
zero sound can be understood as vibration of the Fermi sphere and 
propagates due to the presence of the quasi-particle 
interaction. 

\vskip 0.3cm 
The effective quark theory (\ref{eq:4.64},\ref{eq:4.65}) was studied 
in~\cite{larr97} in great detail employing standard techniques. 
The resulting effective potential was already shown in figure 
\ref{fig:3} and \ref{fig:4}, which were 
used for illustration purposes before. 

\vskip 0.5cm 
{\bf Acknowledgments: } It is a pleasure to thank my collaborators 
Mannque Rho and Hugo Reinhardt. I thank Niels Schopohl for interesting 
discussions on Fermi liquids, and Michael Baake for helpful 
information in the same context. I am indebted to H.~Reinhardt for support.

\end{document}